# Carrier providers or carrier killers: the case of Cu defects in CdTe solar cells

Ji-Hui Yang,[1,*] Wyatt K. Metzger,[1] and Su-Huai Wei[2,*]

[1]National Renewable Energy Laboratory, Golden, CO 80401, USA

[2]Beijing Computational Science Research Center, Beijing 100094, China

## Abstract

Defects play important roles in semiconductors for optoelectronic applications. Common intuition is that defects with shallow levels act as carrier providers and defects with deep levels are carrier killers. Here, taking the Cu defects in CdTe as an example, we show that shallow defects can play both roles. Using first-principles calculation methods combined with thermodynamic simulations, we study the dialectic effects of Cu-related defects on hole density and lifetime in bulk CdTe over a wide range of Cu incorporation conditions. Because $Cu_{Cd}$ can form a relatively shallow acceptor in CdTe, we find that increased Cu incorporation into CdTe indeed can help achieve high hole density; however, too much Cu can cause significant non-radiative recombination. We discuss two strategies to balance the contradictory effects of Cu defects based on the calculated impact of Cd chemical potential, copper defect concentrations, and incorporation temperature on lifetime and hole density. The results indicates that to optimize the Cu doping in CdTe, it is important to control the total amount of Cu incorporated into CdTe and optimize the match between the Cu incorporation temperature and the Cd chemical potential. These findings can help understand the roles of Cu defects in CdTe and the potential complex defect behaviors of relatively shallow defect states in semiconductors for optoelectronic applications.

Defects are known to play important roles in determining the electronic properties of semiconductors, especially those for optoelectronic applications such as photovoltaics (PV).[1,2] Generally speaking, there are two important kinds of defects. On one hand, defects with shallow defect transition energy levels can provide free carriers by ionization; thus, they determine the carrier densities and Fermi levels, which contribute to the solar cell open-circuit voltage ($V_{oc}$) and fill factor.[3,4] A larger amount of such defects will shift the Fermi level toward the band edges, enabling larger $V_{oc}$. On the other hand, defects especially those with deep transition energy levels can cause the non-radiative Shockley-Read-Hall (SRH) recombination of photo-generated electrons and holes, thereby decreasing photo-current, fill factor, and $V_{oc}$.[5] In general, these two kinds of defects are different and a common approach to improve the performance of a solar cell is to increase the carrier providers and decrease the lifetime killers. However, as the defect concentration increases, it will contribute more to recombination, as seen from the relation between the SRH carrier lifetime and defect densities, i.e., $\tau = (B_D N_D)^{-1}$, where $\tau$ is the carrier lifetime, $B_D$ is the defect carrier capture rate, and $N_D$ is the defect density. In this case, a contradiction arises that an increased density of defects with shallow defect levels not only can drive the Fermi level to the band edges and thus help increase the $V_{oc}$ and performance but also might enhance the non-radiative recombination and decrease the photo-current and $V_{oc}$ and overall performance. This contradictory behavior can be more prominent in II-VI materials because the acceptors usually have defect transition energy levels of 0.1 eV above the valence band maximum (VBM) and carrier compensation is common. When there is strong compensation between donor and acceptor defects, more defects have to be created to achieve a certain carrier concentration. In some cases, the defect concentration can be several orders of magnitude greater than the carrier concentration. When this happens, the non-radiative recombination induced by defects even with relatively shallow defect levels can be significant and limit the device performance. Consequently, optimization must balance the advantageous and deleterious impacts of these defects. Here, we show that, Cu doping in CdTe is such a situation.

As one of the leading thin-film solar cell materials for low-cost and high-efficiency PV applications, CdTe has been extensively studied during the past half century.[6,7] Recently, First Solar has achieved a world-record cell efficiency of 22.1%.[8] Experimentally, it is well known that Cu is crucial to attain a relatively high hole concentration and thus a high energy conversion efficiency. However, the exact roles of Cu in enhancing and even limiting CdTe solar cell efficiency are not yet well understood. It has been reported that Cu can reduce the back contact barrier,[9,10] enhance *p*-type doping in CdTe and modify carrier lifetime.[11-13] Recent experiments also show that Cu-related defects can cause non-radiative recombination of photo-generated carriers.[14,15] In addition, because Cu is a mobile species, its concentration and defect nature can change during operation, creating metastability and degradation effects.[16,17] Due to the important roles that Cu plays in CdTe solar cells, it is of great interest to clarify whether and how Cu-related defects can act as both carrier providers and carrier killers. Answering such questions also provides insight on the dialectic roles of somewhat shallow defects in semiconductors.

In this work, we use first-principles calculation methods combined with thermodynamic simulations to examine the impact of Cu-related defects in bulk CdTe. The calculations are performed using density-functional theory (DFT) as implemented in the VASP code[18,19] with detailed methodology described in Ref. 20. The calculated chemical potential conditions are: $\mu_{Cd} + \mu_{Te} = \Delta H_f(CdTe) = -1.17$ eV and $\mu_{Cu} + \mu_{Te} \leq \Delta H_f(CuTe) = -0.25\ eV$, where $\Delta H_f(CdTe)$ and $\Delta H_f(CuTe)$ are the formation enthalpies of perfect CdTe and CuTe bulks, respectively. For simplicity and practical reason, here we focus on Cu on Cd sites and Cu interstitials because other Cu-related defects and intrinsic CdTe defects are less dominant.[20]

The calculated defect formation energies of $Cu_{Cd}$ and $Cu_i$ as functions of Fermi levels are shown in Fig. 1, where two chemical potential conditions are considered. In Fig. 1(a) for an extremely Cd-poor condition ($u_{Cd}$ = -1.17), $Cu_{Cd}$ always has smaller formation energy than that of $Cu_i$, so their compensation is very weak. Fig. 1(b) indicates that in a slightly Cd-rich condition ($u_{Cd}$ = -0.57), $Cu_{Cd}$ can be strongly compensated by $Cu_i$ when the Fermi level is close to the crossing point of their formation energy lines. The (0/-) transition energy level of Cu$_{Cd}$ is calculated to be 0.16 eV above the VBM of CdTe using Heyd-Scuseria-Ernzerhof (HSE06) hybrid functionals,[21] in good agreement with experimental measurements.[15] The (0/+) transition energy level of Cu$_i$ is 0.15 eV below the conduction band minimum (CBM) of CdTe.

Although these defect levels are close to band edges, their in-gap positions indicate that they can still be involved in non-radiative recombination. Here, we have quantitatively calculated the non-radiative recombination rates of these two defects using the recently developed method that takes into account multiphonon emissions under static approximations.[22,23] This method has shown very consistent results with experimental results on II-VI and III-V semiconductors.[23,24] For the electron trapping rate, our calculated results at T=300 K are $2.55 \times 10^{-10} cm^3 s^{-1}$ and $9.96 \times 10^{-9} cm^3 s^{-1}$ for Cu$_{Cd}$ and Cu$_i$, respectively. For the hole trapping rate, the results are $2.84 \times 10^{-8} cm^3 s^{-1}$ and $6.82 \times 10^{-10} cm^3 s^{-1}$ for Cu$_{Cd}$ and Cu$_i$, respectively. Although it is easy for Cu$_{Cd}$ to trap holes because of its defect level being close to the VBM, Cu$_i$ can easily trap electrons because its level is close to the CBM. Note that, the carrier trapping process is usually limited by the slow step. Considering the electron and hole densities are comparable under sunlight in CdTe,[24] the trapping rate due to Cu$_{Cd}$ is determined by the electron trapping process and the trapping rate due to Cu$_i$ is determined by the hole trapping process. Correspondingly, the SRH lifetimes due to Cu$_{Cd}$ and Cu$_i$ are $\tau_{Cu_{Cd}} = (B_n N_{Cu_{Cd}})^{-1}$ and $\tau_{Cu_i} = (B_p N_{Cu_i})^{-1}$, with $B_n$ and $B_p$ being the electron and hole trapping rates, respectively. Compared to the trapping rate of the dominant intrinsic recombination center Te$_{Cd}^{2+}$,[24] the trapping rates of Cu$_{Cd}$ and Cu$_i$ are two or three orders of magnitude smaller, i.e., $10^{-7} cm^3 s^{-1}$ for Te$_{Cd}^{2+}$ versus $10^{-10} cm^3 s^{-1}$ for these Cu defects. Despite this, Cu defects can still cause significant recombination if the concentration of Cu incorporated in CdTe is high. For example, if the total concentration of Cu defects is larger than $10^{17} cm^{-3}$, the SRH carrier lifetime τ will be smaller than tens of nanoseconds determined from $\frac{1}{\tau} =$ $\frac{1}{\tau_{Cu_{Cd}}} + \frac{1}{\tau_{Cu_i}}$ and further to several nanoseconds if Cu concentration exceeds $10^{18}$ cm$^{-3}$.

Consequently, a contradictory consideration of hole density and carrier lifetime arises: from the perspective of obtaining a high hole density (i.e., $\geq 10^{16}$ cm$^{-3}$), a large Cu incorporation is preferred; from the perspective of obtaining a high carrier lifetime, less Cu incorporation is preferred. In practice, the incorporation of Cu into CdTe can be controlled by the incorporation or annealing temperatures and the Cd chemical potentials as well as other factors such as diffusion times. To get a high hole density, the extremely Cd-poor condition and a high incorporation temperature are beneficial because more Cu$_{Cd}$ defects can be created. To get a high carrier lifetime and avoid excessive Cu incorporation, the not-so-poor Cd condition and a low incorporation temperature are helpful.

Due to the complex roles of the annealing temperature and the Cd chemical potential in determining the optimal Cu incorporation conditions in CdTe, we perform thermodynamic simulations. First, we consider the case when Cu is incorporated into CdTe under thermodynamic equilibrium growth. Under this condition and within the dilute limit, the density of a defect $\alpha$ with charge state $q$ is assumed to reach equilibrium at the incorporation temperature and can be determined from:

$$n(\alpha, q) = N_{site} g_q e^{-\Delta H_f(\alpha,q)/k_B T}, \quad (1)$$

where, $N_{site}$ is the number of possible sites per volume for defect $\alpha$, $g_q$ is the degeneracy factor,[25,26] and $\Delta H_f(\alpha, q)$ is the defect formation energy[27] defined as:

$$\Delta H_f(\alpha, q) = E(\alpha, q) - E(host) + \sum_i n_i(E_i + \mu_i) + q[\varepsilon_{VBM}(host) + E_F], \quad (2)$$

which is a function of chemical potentials $\mu_i$ of involved elements and $E_F$. At a given temperature, the thermally excited electron density $n_0$ and hole density $p_0$ are also functions of Fermi level, which are given as:

$$n_0 = N_c e^{-\frac{E_c - E_F}{k_B T}}, N_c = 2\frac{(2\pi m_n^* k_B T)^{\frac{3}{2}}}{h^3},$$
$$p_0 = N_v e^{-\frac{E_F - E_v}{k_B T}}, N_v = 2\frac{(2\pi m_p^* k_B T)^{3/2}}{h^3}. \quad (3)$$

Here, $N_c$ is the temperature-dependent effective density of states (DOS) of the conduction band and $N_v$ is the effective DOS of the valence band. $m_n^*$ (0.095 $m_0$ for CdTe) and $m_p^*$ (0.84 $m_0$ for CdTe) are effective masses of electrons and holes.[28] The charge neutralization condition for Cu doped CdTe requires that:

$$p_0 + n(Cu_i^+) = n_0 + n(Cu_{Cd}^-), \quad (4)$$

By solving Eqs. (1)-(4) self-consistently, we can obtain the Cu defect densities in CdTe at given chemical potentials, as well as carrier densities and Fermi levels assuming everything reaches equilibrium at a given incorporation temperature. In reality, the Cu incorporation temperature is often well above the room temperature and then the system is cooled down to 300 K. During the annealing process, it's reasonable to assume that if the cooling rate is high, the amount of Cu at substitutional sites and the amount of Cu interstitials are kept the same as they are at the high incorporation temperature,[29] although their charge states might be redistributed following the Boltzmann distributions. For example, $Cu_{Cd}$ defect can be redistributed between its neutral and -1 states by:

$$n(Cu_{Cd}^0) = N_{Cu_{Cd}} \times \frac{g_0 e^{-\Delta H_f(Cu_{Cd}^0)/k_B T}}{g_0 e^{-\Delta H_f(Cu_{Cd}^0)/k_B T} + g_{-1} e^{-\Delta H_f(Cu_{Cd}^-)/k_B T}};$$

$$n(Cu_{Cd}^-) = N_{Cu_{Cd}} \times \frac{g_0 e^{-\Delta H_f(Cu_{Cd}^-)/k_B T}}{g_0 e^{-\Delta H_f(Cu_{Cd}^0)/k_B T} + g_{-1} e^{-\Delta H_f(Cu_{Cd}^-)/k_B T}}, \quad (5)$$

where $N_{Cu_{Cd}}$ is the total density of Cu at Cd substitutional sites in CdTe calculated at the high incorporation temperature; $g_0$ and $g_{-1}$ are the degeneracy factors, which are 4 and 1 for $Cu_{Cd}^0$ and $Cu_{Cd}^-$, respectively; $\Delta H_f(Cu_{Cd}^0)$, and $\Delta H_f(Cu_{Cd}^-)$ are the defect formation energies. By solving Eqs. (2)-(5) at the room temperature using the defect information at the high incorporation temperature, we can obtain the room-temperature Cu defect densities in CdTe as well as carrier densities and Fermi levels at given chemical potentials.

We perform simulations for the cases when Cu is incorporated into CdTe at extremely Cd-poor and slightly Cd-rich conditions over a wide range of annealing temperatures. After quickly quenching to the room temperature, the obtained hole density, Fermi level, defect density and carrier lifetime are shown in Fig. 2. At given Cd chemical potential conditions, the incorporated Cu concentration into CdTe monotonically increases with the increased annealing temperature [Figs. 2(b) and 2(e)] as expected because more defects can be created at a higher temperature. At the same time, the hole density monotonically increases and the Fermi level monotonically decreases at extremely Cd-poor conditions when the annealing temperature increases [Figs. 2(a)]. This trend occurs because the band-edge thermal excitation at finite temperatures always tends to drag the Fermi level to the middle of the bandgap, where $Cu_{Cd}$ always has lower formation energy than $Cu_i$. As a result, more $Cu_{Cd}$ will be created than $Cu_i$ with increased annealing temperature. After annealing, although part of $Cu_{Cd}$ defects will convert from negatively charged states to neutral states, the net density difference between $Cu_{Cd}^-$ and $Cu_i^+$ still increases with increased annealing temperature, resulting in increased hole density and decreased Fermi level. Results at slightly Cd-rich conditions are a little different. In this case, the formation energy difference between $Cu_{Cd}$ and $Cu_i$ is smaller. Although more $Cu_{Cd}$ will be created than $Cu_i$ at the higher annealing temperature, their density difference decreases. After annealing, partial conversion of $Cu_{Cd}^-$ defects into its neutral states further reduces the net density difference between $Cu_{Cd}^-$ and $Cu_i^+$. The consequence is that there is a peak of hole density and a dip of Fermi level when the annealing temperature is about 1000 K in Fig. 2(d). At a given annealing temperature, the incorporated Cu concentration and the resulting hole density decrease when Cd chemical potential increases from the extremely Cd-poor condition to the slightly Cd-rich condition. This is because the formation energy of $Cu_{Cd}$ increases, and thus less $Cu_{Cd}$ can be created to contribute to the hole density when the Cd chemical potential increases. Similarly, the Fermi level increases with the increase of Cd chemical potential.

Our simulations confirm that the higher incorporation temperature and the extremely Cd-poor condition are indeed helpful to get higher hole density. However, the incorporated Cu concentration can be too excessive and cause significant non-radiative recombination. For example, with the incorporation temperature T=800 K and under extremely Cd-poor condition, $\sim 10^{18} \text{cm}^{-3}$ Cu can be incorporated into CdTe and the hole density can reach $\sim 10^{17} \text{cm}^{-3}$. However, the carrier lifetime meanwhile is limited to several nanoseconds. Therefore, in practice, one needs to balance the hole density and the carrier lifetime by optimally matching the incorporation temperature and the Cd chemical potential. Experimentally, it's known that a high hole density above $10^{16} \text{cm}^{-3}$ with a carrier lifetime of tens of nanoseconds is sufficient to approach $V_{oc}$ of 1 V.[30] In the case of extremely Cd-poor conditions, the optimal incorporation temperature would be around 600─700 K according to our simulations [Fig. 2(c)]. Similarly, in the case of slightly Cd-rich conditions, the optimal incorporation temperature would be around 800─900 K [Fig. 2(f)]. Contrary to intuition, a higher incorporation temperature and extremely Cd-poor condition are actually not necessary to realize the optimal Cu incorporation. Using the thermodynamic simulation methods, we estimate the incorporation temperatures and Cd chemical potentials that can enable both a high hole density above $10^{16} \text{cm}^{-3}$ and a carrier lifetime of tens of nanoseconds in Fig. 3. Generally speaking, a lower incorporation temperature is needed with more Cd-poor stoichiometry. Our simulations thus offer a theoretical guide on how to optimally choose the incorporation temperatures and the Cd chemical potentials when Cu is incorporated into CdTe under thermodynamic equilibrium growth conditions.

In practice, an alternative way to control the Cu incorporation into CdTe is to control the total amount of Cu concentration, i.e., by controlling the amount of Cu diffusion sources or the diffusion times. In this case, the incorporated Cu will be distributed into different Cu defects following the Boltzmann distributions similar to Eq. (5) except that now the total Cu density $N_{Cu}$ is distributed into $Cu_{Cd}^0$, $Cu_{Cd}^-$, $Cu_i^0$ and $Cu_i^+$ according to their weights determined from their defect formation energies. Our above simulations indicate that a reasonable Cu concentration in CdTe for PV applications is about $10^{17} \text{cm}^{-3}$, which is sufficient to provide a hole density above $10^{16} \text{cm}^{-3}$ and retain the carrier lifetimes of ten of nanoseconds. More Cu incorporation will reduce the carrier lifetime and less Cu incorporation can't achieve a high hole density.

While this strategy is not necessarily restricted to equilibrium growth condition, the final hole density and carrier lifetime are still determined by the incorporation temperature and Cd chemical potential. To examine this, we perform thermodynamic simulations over a wide range of incorporation temperatures and Cd chemical potentials with the total amount of Cu fixed at $10^{17} \text{cm}^{-3}$. Fig. 4 shows our simulation results. Here, the carrier lifetimes are always larger than 20 nanoseconds due to the limited Cu incorporation. In addition, to achieve a high hole density at the extremely Cd-poor condition, a low incorporation temperature of 300–400 K is sufficient [Fig. 4(b)]. However, with increased Cd chemical potential, a higher incorporation temperature is necessary. For example, under the slightly Cd-rich condition in the case of Fig. 1(b), an incorporation temperature of about 800 K is required to get a hole density above $10^{16} \text{cm}^{-3}$. In fact, for a wide range of Cd chemical potentials, we can often get hole density $\geq 10^{16} \text{cm}^{-3}$ and a carrier lifetime of nearly 40 ns [Fig. 4(c)] as long as the Cu incorporation temperature is $\geq 800$ K and the total Cu concentration is limited to be $10^{17} \text{cm}^{-3}$. Again, our simulations show that a higher incorporation temperature and extremely Cd-poor chemical potential are not necessary to realize the optimal Cu incorporation. Instead, the total Cu concentration in CdTe and a good match between the incorporation temperature and the Cd chemical potentials are important.

In conclusion, we have studied the roles of Cu-related defects in bulk CdTe. In general, opposing trends exist between lifetime and hole density with Cu incorporation. We show that to optimize the Cu doping in CdTe, it is important to control the total amount of Cu incorporated into CdTe and the match between the incorporation temperature and the Cd chemical potentials. The findings help address the role of Cu defects for CdTe solar technology and more generally provide insight into potential dialectic behaviors for somewhat shallow defects in semiconductors for optoelectronic applications.

This work was funded by the U.S. Department of Energy, EERE/SunShot program, under Contract No. DE-AC36-08GO28308. Work at Beijing CSRC is supported by NSAF joint program under Grant Number U1530401. The calculations are done on NREL's Peregrine supercomputer and NERSC's supercomputer.

Email: jihuiyang2016@gmail.com; suhuaiwei@csrc.ac.cn

**Figures:**

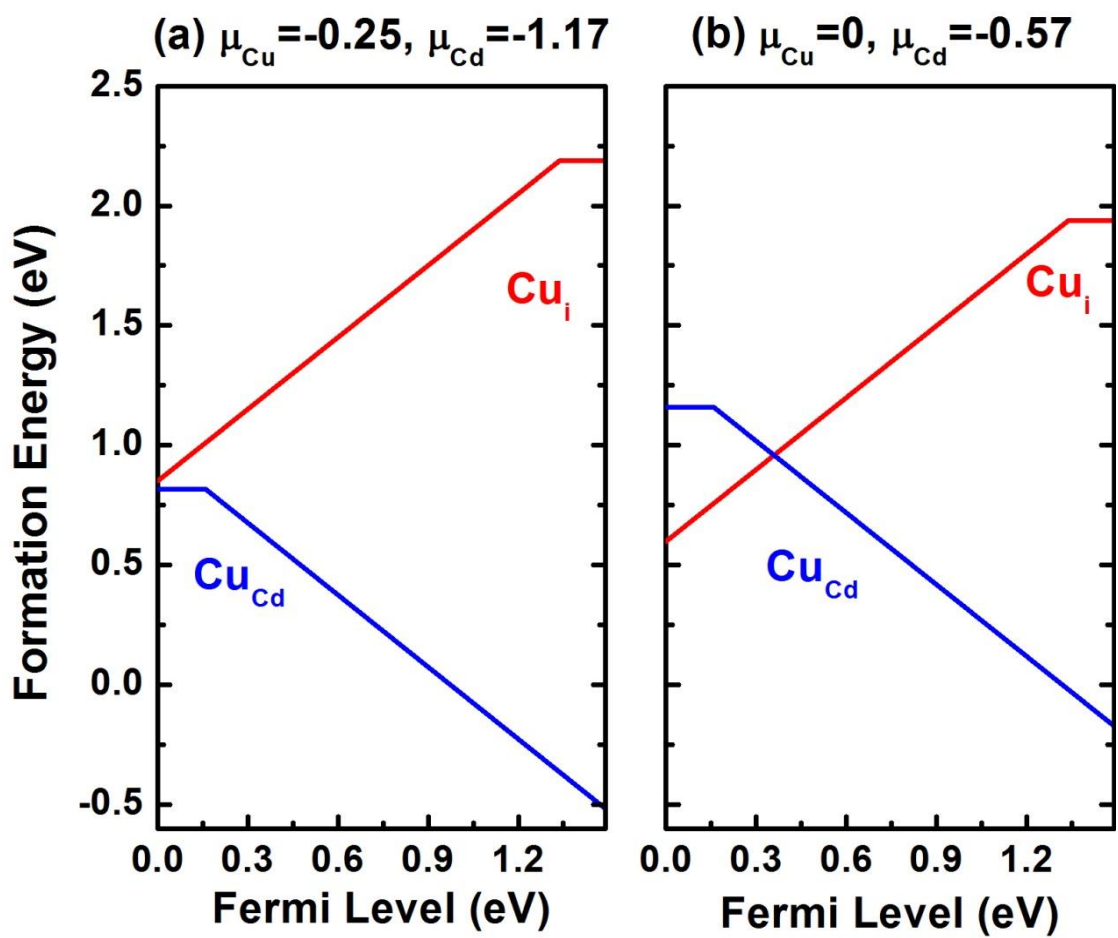

Fig. 1. HSE06 calculated formation energies of $Cu_{Cd}$ and $Cu_i$ as functions of Fermi levels (referenced to the VBM) under (a) extremely Cd-poor condition and (b) slightly Cd-rich condition.

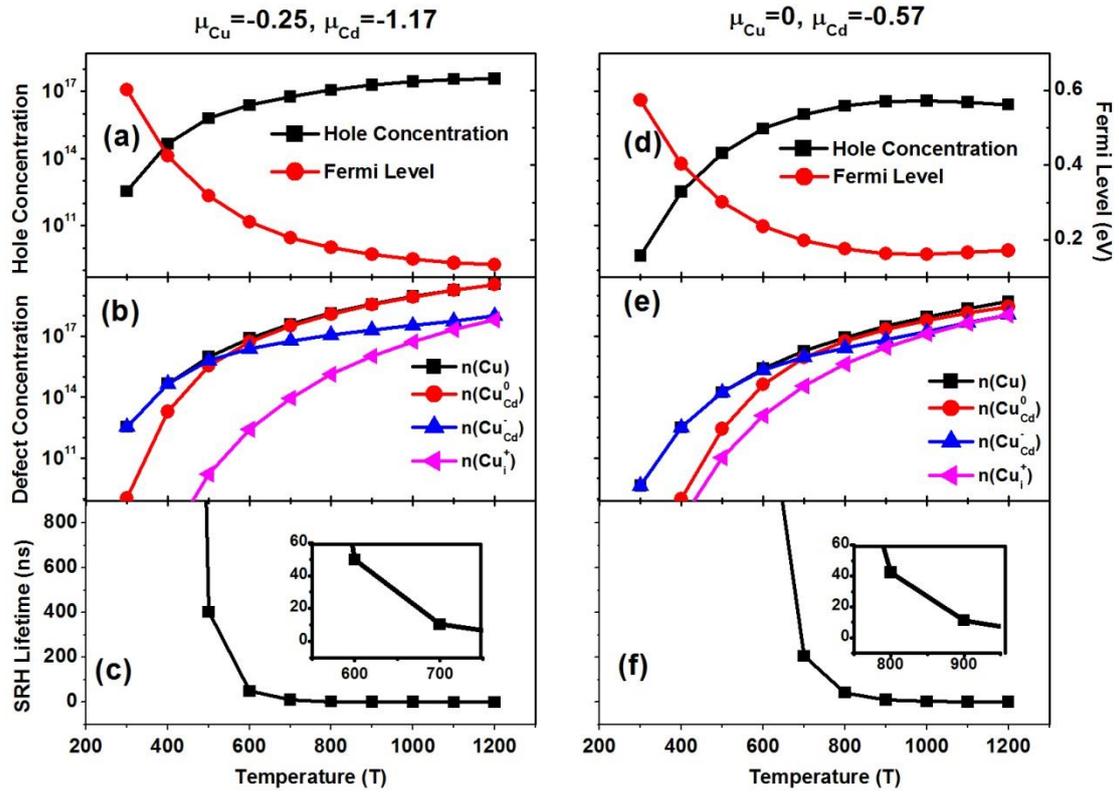

Fig. 2. (a) and (d) Fermi level and hole density, (b) and (e) Cu defect densities, (c) and (f) SRH carrier lifetime in annealed CdTe samples after annealing CdTe from different Cu incorporation temperatures under thermodynamic equilibrium condition to T=300 K for extremely Cd-poor case (left panel) and slightly Cd-rich case (right panel).

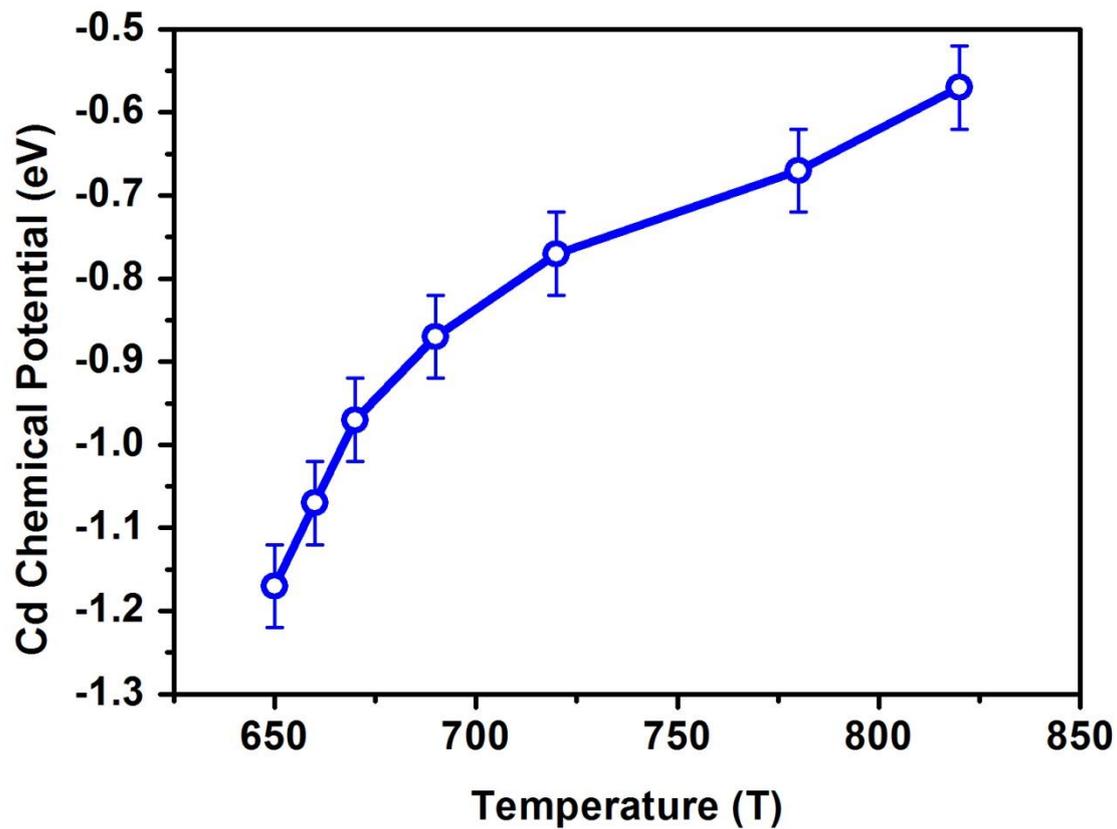

Fig. 3. The estimated optimal match between the Cu incorporation temperature and the Cd chemical potential to achieve both a high hole density above $10^{16}$ cm$^{-3}$ and a carrier lifetime of tens of nanoseconds. The error bars indicate an estimation.

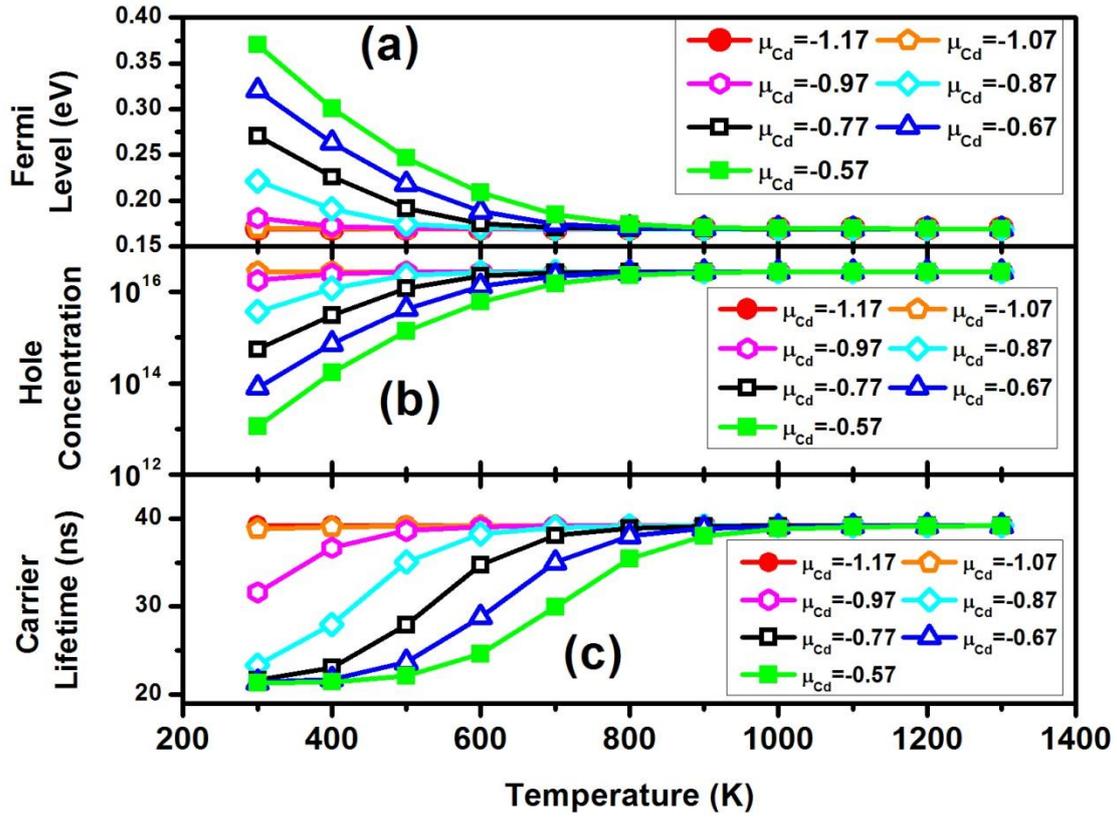

Fig. 4. (a) Fermi level, (b) hole density, and (c) SRH carrier lifetime in annealed CdTe samples after annealing CdTe samples from different Cu incorporation temperatures to T=300 K for different Cd chemical potentials.